# Discriminative sEMG-based features to assess damping ability and interpret activation patterns in lower-limb muscles of ACLR athletes

Mehran Hatamzadeh [a, b, d], Ali Sharifnezhad [c*], Reza Hassannejad [d], Raphael Zory [a, e]

[a] Université Côte d'Azur, LAMHESS, Nice, France.

[b] Université Côte d'Azur, Inria, Sophia Antipolis, France.

[c] Department of Sport Biomechanics and Technology, Sport Sciences Research Institute (SSRI), Tehran, Iran.

[d] Department of Mechanical Engineering, University of Tabriz, Tabriz, Iran.

[e] Institut Universitaire de France (IUF), Paris, France.

**\* Corresponding Author:**

**Ali Sharifnezhad**, Sport Science Research Institute of Iran, No. 3, 5th St., MirEmad St., Motahari Ave., Tehran, Iran. Postal Code:1587958711. Email: a.sharifnezhad@ssrc.ac.ir – ORCID: 0000-0002-5855-1441

**Co-authors:**

**Mehran Hatamzadeh** – Email: mehran.hatamzadeh@inria.fr – ORCID: 0000-0002-5183-4801

hatamzadeh.mehran@gmail.com

**Reza Hassannejad** – Email: hassannejhad@tabrizu.ac.ir – ORCID: 0000-0002-5249-3455

**Raphael Zory** – Email: Raphael.ZORY@univ-cotedazur.fr – ORCID: 0000-0003-3566-3229

**Abstract:**

Objective: The main goal of the athletes who undergo anterior cruciate ligament reconstruction (ACLR) surgery is a successful return-to-sport. At this stage, identifying muscular deficits becomes important. Hence, in this study, three discriminative features based on surface electromyographic signals (sEMG) acquired in a dynamic protocol are introduced to assess the damping ability and interpret activation patterns in lower-limb muscles of ACLR athletes.

Methods: The features include the median frequency of the power spectrum density (PSD), the relative percentage of the equivalent damping or equivalent stiffness derived from the median frequency, and the energy of the signals in the time-frequency plane of the pseudo-Wigner-Ville distribution (PWVD). To evaluate the features, 11 healthy and 11 ACLR athletes (6 months post-reconstruction surgery) were recruited to acquire the sEMG signals from the medial and the lateral parts of the hamstrings, quadriceps, and gastrocnemius muscles in pre- and post-fatigue single-leg landings.

Results: A significant damping deficiency is observed in the hamstring muscles of ACLR athletes by evaluating the proposed features. This deficiency indicates that more attention should be paid to this muscle of ACLR athletes in pre-return-to-sport rehabilitations.

Conclusion: The quality of electromyography-based pre-return-to-sport assessments on ACLR subjects depends on the sEMG acquisition protocol, as well as the type and nature of the extracted features. Hence, combinatorial application of both energy-based features (derived from the PWVD) and power-based features (derived from the PSD) could facilitate the assessment process by providing additional biomechanical information regarding the behavior of the muscles surrounding the knee.

# 1. Introduction

Non-contact injury to the anterior cruciate ligament (ACL) is a prevalent sports-related injury with long recovery times [1,2]. Sudden changes in the movement direction, muscular weakness, and instability of the joints are some of the main reasons for this injury [3,4]. Some secondary reasons may also increase the probability of this injury occurring, including muscular fatigue. Muscular fatigue increases the risk of lower-limb injuries, especially ACL injury, by altering neuromuscular function and biomechanical patterns [5,6]. Athletes suffering from this injury usually undergo ACL reconstruction (ACLR) surgery to be able to return-to-sport [7,8]. Nevertheless, even reconstruction cannot guarantee a successful return-to-sport and, after reconstruction, while roughly 81% of the patients become able to participate in sports, only 65% return to their pre-injury physical readiness, and only 55% regain the ability to compete in high-demand sports [7,9]. One of the most commonly reported consequences after ACL reconstruction is asymmetry and deficit in muscles' strength [10,11]. These consequences following ACL reconstruction cause ACLR subjects to be at greater risk of ACL re-rupture, meniscus injuries, and early development of osteoarthritis [1,12]. In this regard, statistics show that out of every 4 ACLR athletes, one suffers ACL re-rupture after returning to sport, which highlights the importance of pre-return-to-sport assessments and rehabilitation [13, 14]. The expected return-to-sport time is 6 to 12 months post-surgery, before which time functional impairments should be identified and improved to ensure a more successful return-to-sport [15]. Usually, analyzing movement in ACLR athletes who are at the stage of returning to sport, is done using dynamic protocols involving rapid decelerations, a period of stabilization, and a phase of ground contact such as landing tests [7,16,17].

One of the most common methods of assessing ACL health status is magnetic resonance imaging (MRI) [18,19]. However, this equipment is not only expensive but also imposes high costs on the patient. For this reason, efforts are focused on making these assessments less expensive, while maintaining an acceptable accuracy [20]. One of the less expensive and non-invasive alternative tools is surface electromyography (sEMG). Since ACL injury and its reconstruction affects muscular activity patterns, sEMG can be used as a monitoring tool to assess ACL health status based on surrounding muscles' activity or monitoring its recovery following ACL reconstruction [21-23]. It should be noted that the sEMG itself does not provide useful information, but rather the features extracted from the acquired signals give meaning and value to the sEMG [24]. Therefore, the type and nature of the extracted features are of special importance and can directly affect the quality of the assessment. In this regard, many sEMG-based approaches have been developed to assess ACL health status, to identify movement asymmetries post-reconstruction surgery, to study knee joint mechanics and loadings on its ligaments during variety of tasks [25-28]. For instance, a comparison of two methods of diagnosing ACL health status using artificial intelligence-based algorithms shows that selecting appropriate features from the raw sEMG signal can not only provide more useful information but also makes the diagnosis with the use of neural networks easier, hence increasing the final accuracy [29-31]. However, literature review indicates that there is still lack of reliable and valid sEMG-based features to be used, especially in return-to-sport phase following ACL reconstruction surgery [32]. It is worth mentioning that muscular activity is the sum of the activity of all motor units within a muscle and, to better interpret the units' functioning using sEMG, combined time-frequency analysis is usually recommended [33,34]. One such analysis is the median frequency (MF) of electromyographic signals, based on which utilization and activation patterns of muscular units can be determined [35]. It has been shown that the MF of the electromyographic signals could be used as an index to identify recruitment strategies of muscular units, to infer the conduction velocity within the muscle fibers, and the amount of force produced by muscles [33,36,37]. In this regard, Drechsler and her colleagues assessed variations of the MF of the rectus femoris muscle's sEMG signal acquired during maximal voluntary contractions (MVC) at 1 and 3 months after ACL reconstruction surgery [33]. However, assessing MF when the athlete is on the verge of return-to-sport, in other muscles around the knee, or when performing dynamic protocols has not been addressed in that research. Recently, it has been shown that time-frequency analysis of the sEMG signals from the energy point of view, using the pseudo-Wigner-Ville distribution, can reveal behavioral distinctions between the muscles of healthy and ACLR athletes and is a tool to monitor muscles' ability to damp the energy of the sEMG signal produced when landing [29,30].

Therefore, this research aims to introduce, extract and evaluate some sEMG-based features that, by providing additional information regarding the behavior of ACLR athletes' muscles when performing a dynamic protocol, could be used in pre-return-to-sport assessments following ACL reconstruction and monitoring ACL health status. Hence, three discriminative features are introduced that enable recruitment patterns of motor units in the muscles to be identified, the effectiveness of rehabilitation programs on muscles to be assessed, and the ability of muscles in damping vibrations and energy moderating to be evaluated.

## 2. Methodology

### 2.1. Subjects

In this research, 22 young male athletes participated as the subjects. According to the literature, at least 10 subjects in each of the healthy and ACLR groups is needed to achieve the test power of 80% in statistical analysis, at the significant level of 5% [38]. On this basis, 11 healthy athletes (age: 23.8±2.8 years, weight: 72.6±7.4 kg, height: 173.1±9.1 cm) and 11 ACLR athletes (age: 24.2±3.7 years, weight: 73.5±7.5 kg, height: 174.2±9.7 cm) were recruited, all of whom had moderate-to-high levels of sports activity with at least 1.5 hours of exercise per day and at least 3 days per week. Approximately 6 months had elapsed from the time of surgery for the ACLR athletes and they were on the verge of returning to sport. Reconstructions were performed using the graft types of the patellar tendon (n=4), semitendinosus/gracilis (n=4), and allograft (n=3). Following the ethical principles of human experimentation described in the Helsinki Declaration, the subjects were selected with the oversight of an orthopedic surgeon and participated in this research after giving informed written consent.

### 2.2. Data acquisition

To acquire the data, first, the correct implementation of the single-leg landing protocol was taught to the athletes and all participants performed practice trials for familiarization. The implementation procedure of this protocol is in such a way that the subject stands on top of a box, putting the non-target foot on the box and holding the target foot slightly ahead of the box. Then, the subject falls down vertically and lands on the target foot, while maintaining their stability for a few seconds on that limb. The target foot was considered as the reconstructed leg of the ACLR athletes and the dominant leg of the healthy athletes [29]. In this process, a 40 cm box was used for both practice trials and the main landing tests [29]. After practice trials, the skin was shaved and cleaned. Following the skin preparation, the adhesive sEMG electrodes were installed on the most prominent bulge of the muscles in parallel orientation with the muscle fibers according to the SENIAM guidelines (Surface Electromyography for the Non-invasive Assessments of Muscles). As is illustrated in Figure (1), two measuring electrodes (+ and -) were installed on each muscle and their corresponding ground/reference electrode was installed on the shank bone or the S2 lumbar vertebra and their cables were fixed to the skin using adhesive tape. The sEMG signals were recorded using disposable Ag/AgCL adhesive electrodes (2 cm inter-electrode distance and 2 cm diameter) and a portable 16-channel ME6000 device (Megawin, MEGA electronics Ltd., Finland) with a sampling frequency of 2000 Hz. Following the sEMG electrodes installation, the maximal voluntary contraction (MVC) of the lower-limb muscles of each subject was recorded. Then, pre-fatigue muscular activities were recorded in three successful tries of single-leg landing for each subject. Immediately after, implementation of the fatigue protocol began. To induce fatigue, each subject performed several sets of two-leg squats with 10 squats per set, in such a way that the flexion angle of the knee was approximately 90 degrees, hands were in parallel with the ground, and 3 single-leg landings were performed after each squat set [21,38]. When implementing the fatigue protocol, continuous verbal feedback was given to the subjects on correct implementation and the knee flexion angle. Performing the fatigue protocol continued until the athlete reached maximum fatigue in the lower limb, assumed to occur when they became unable to perform 5 consecutive squats correctly. Immediately after reaching maximum fatigue, muscular activities in 3 post-fatigue landings were recorded for each subject. All the sEMG signals were recorded with 5 seconds length in which the recording started 2 seconds before the landing [29]. The muscles that were tested included: Medial Hamstring (MH), Lateral Hamstring (LH), Medial Gastrocnemius (MG), Lateral Gastrocnemius (LG), Vastus Medialis (VM), and Vastus Lateralis (VL). In addition to sEMG recording, to detect the time of ground contact, an AMTI force plate (Advanced Medical Technology, Inc., USA) with dimensions 50×50 cm, sampling at 2000 Hz, was used.

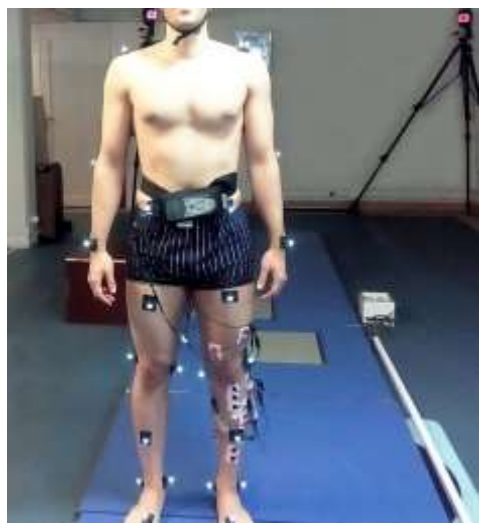

**Figure 1.** An image of a subject with the sEMG electrodes installed over the left leg muscles' bellies. Red electrodes are the measuring electrodes and the black electrodes are the ground/reference electrodes.

## 2.3. EMG Onset Detection

EMG onset time is the moment that the first action potential is produced in the motor units and can be determined from the acquired muscle activity signal [39]. For this purpose, a combination of the Teager–Kaiser energy operator (TKEO) and a threshold-based method was used. It has been shown that this combination increases the accuracy of detecting the muscular activity onset time and has a low computational error [39,40]. As shown in Figure (2), to combine the two algorithms, facilitate onset detection, and to reduce the noise and movement artifacts, the first step is to apply a 6th-order Butterworth band-pass filter on each sEMG signal, with high- and low-pass cutoff frequencies of 30 and 300 Hz, respectively. Afterward, the TKEO was calculated, using the following formula for discrete signals:

$$TKEO(T) = x^2(T) - (x(T-1)x(T+1)) \tag{1}$$

where $x(T)$ represents the discrete sEMG signal's value for the $T^{th}$ sample, $TKEO(T)$ is the TKEO value for the $T^{th}$ sample. Following TKEO calculation, the next steps are rectifying and then applying a $2^{nd}$ order low-pass Butterworth filter at 50 Hz. Then, the sEMG onset time is detected using the threshold-based method as follows:

$$Threshold = \mu + h\sigma \tag{2}$$

where $\mu$ is the mean, $\sigma$ is the standard deviation of the baseline, and $h$ is a preset value that is set to be 15 [39,40]. From the whole 5 second length sEMG signals that starts 2 seconds prior to landing, the first second is used to calculate the mean and standard deviation of the baseline. In the threshold-based method, the onset time is the first sample in the whole signal for which the threshold value in 25 consecutive samples exceeds the threshold value of the baseline [39,40].

After detection of the sEMG onset time, to extract power-based and energy-based features from the sEMG signals, first, all the raw sEMG and MVC signals were band-pass filtered using a 4th order Butterworth filter with high- and low-pass cutoff frequencies of 10 and 500 Hz, respectively [21]. Then, each sEMG signal was normalized with the absolute maximum of its corresponding MVC and was cut from the detected onset time to the end of the reactive phase (250 milliseconds after ground contact) and only the normalized signal within this interval was used to extract the features.

## 2.4. Pseudo Wigner–Ville Distribution

To calculate the energy distribution of the signals from the activity onset time to the end of the reactive phase, the pseudo-Wigner-Ville distribution (PWVD) was used [29,30,41]. This analysis can illustrate time-frequency characteristics of non-stationary signals and represent an energy distribution at different time-frequency intervals [29,30]. As shown in Figure (2), the computational procedure of the PWVD is such that, first, to avoid aliasing phenomena, the signal should be transformed into an analytical form using the Hilbert transform [41]. Then the Wigner-Ville distribution (WVD) and subsequently the PWVD are calculated as follows:

$$\begin{aligned} &x_a(n) = x(n) + jH(x(n)) \\ &WVD(l,k) = \frac{1}{N}\sum_{n=0}^{N-1} x_a(l+n) x_a^*(l-n) e^{-j(\frac{4\pi}{N})nk}, \quad k = 0,1,2,...,N-1 \\ &PWVD = WVD \otimes GW \end{aligned} \quad (3)$$

Where $x(n)$ is the real sEMG signal value at sample $n$, $H(x(n))$ is the Hilbert transform of the signal, $x_a$ is the analytic form of the signal, $x_a^*$ is the complex conjugate of the analytic signal, $N$ is the sEMG signal's length, and $l$ and $k$ are the discrete time and frequency respectively. The PWVD is the result of a two-dimensional convolution of the signal's WVD and a 2D Gaussian window ($GW$), which is then normalized by the peak value of the distribution [29,30,41]. Examples of the normalized sEMG signals of the lateral hamstring muscle (LH) and their corresponding PWVDs in pre- and post-fatigue landings of a healthy athlete and an ACLR athlete are illustrated in Figure (3). It is worth mentioning that the total volume under the PWVD represents the signal's total energy over all time and all frequencies and can be divided into 2 different phases [29,30]. The first phase is from the moment of sEMG onset to the moment of initial ground contact, called the preparation phase [29,30]. The second phase is from the moment of initial ground contact to 250 milliseconds later, called the reactive activity phase or reactive muscular firing phase [21,42,43]. Hence, to compare the energy of the sEMG signals, the volume underneath the PWVD in the reactive phase is calculated and expressed as a percentage of the total volume under the distribution.

## 2.5. Median Frequency of Power Spectrum Density

Power spectrum density analysis is used to calculate the median frequency (MF) of the sEMG signals. This analysis transforms the signal in the time domain to the frequency domain and provides the power of the signal at each frequency based on the Fourier transform of the autocorrelation function of the input signal. After calculating the power at each frequency, the median frequency is the frequency that divides the power spectrum density into two areas with equal power [33]. In this way, the median frequency of lower-limb sEMG signals in both pre- and post-fatigue states was calculated and assessed.

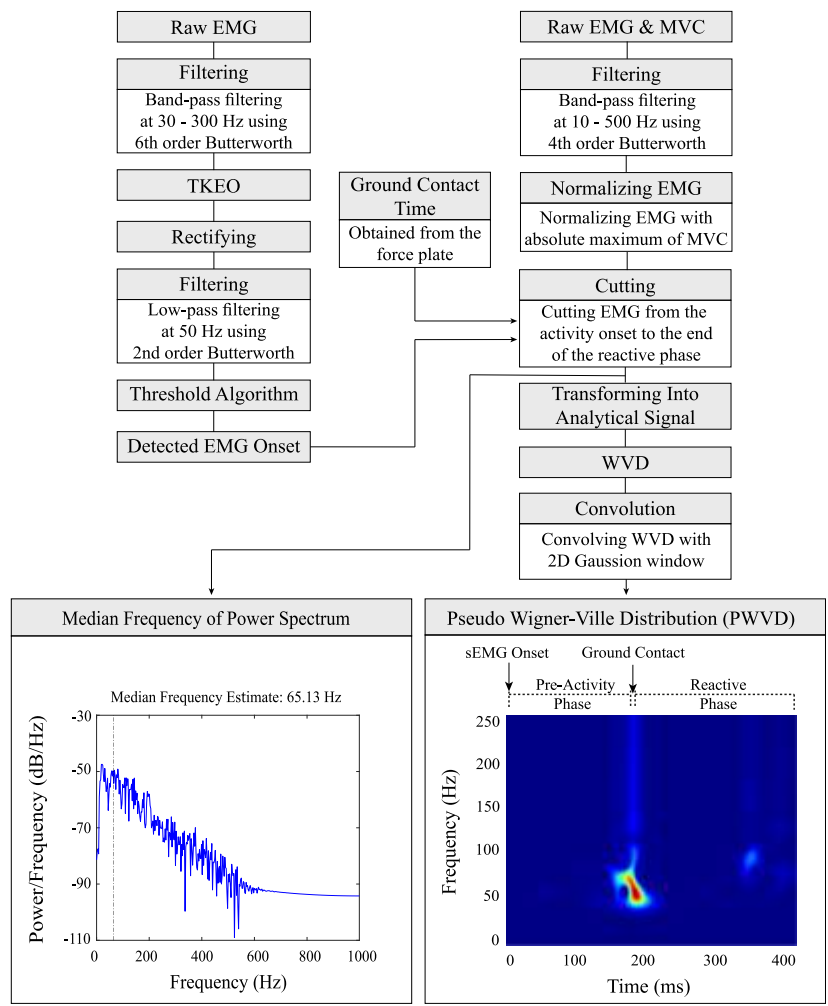

**Figure 2.** Computational steps for the sEMG onset detection, calculation of the pseudo-Wigner-Ville distribution, and median frequency of the power spectrum density.

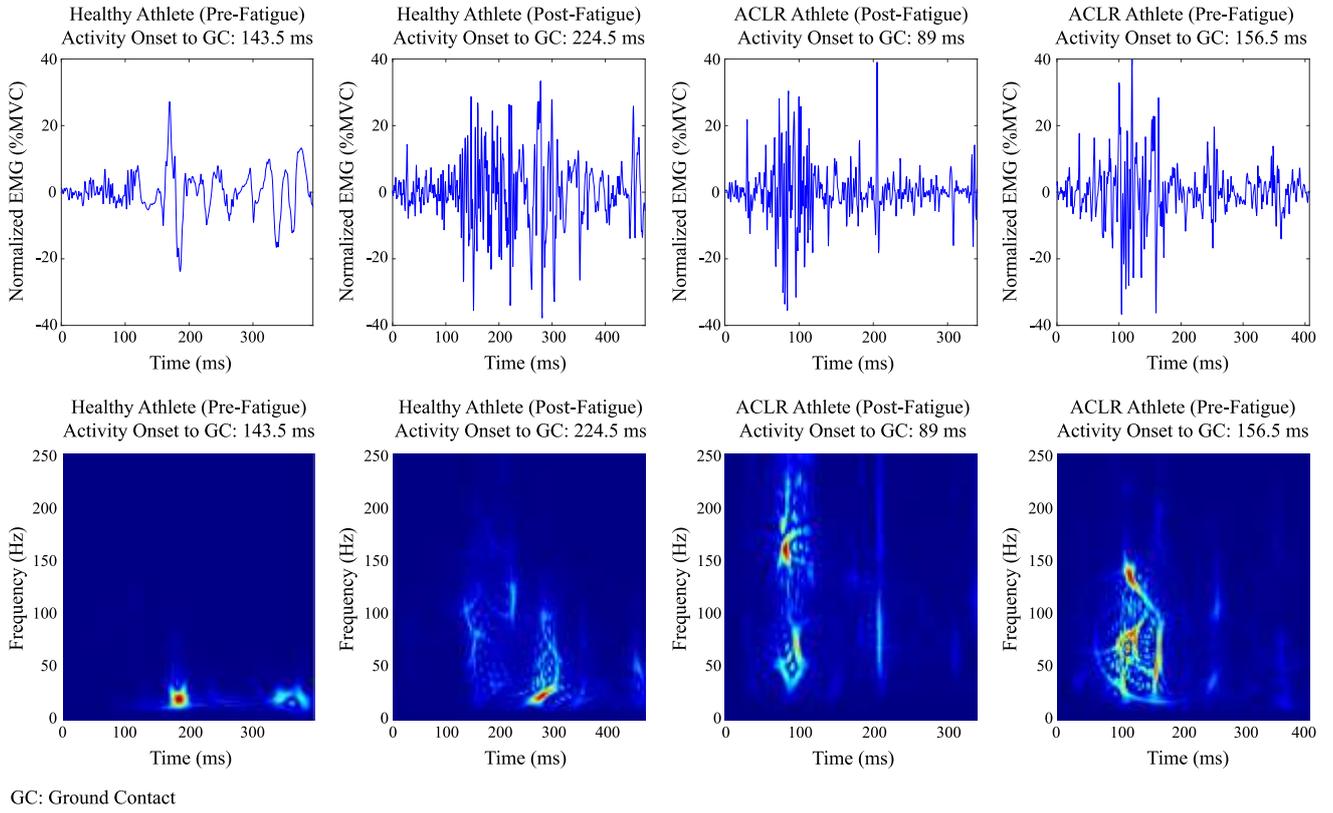

GC: Ground Contact

**Figure 3.** Examples of the normalized sEMG signals of the lateral hamstring (LH) muscle and their corresponding PWVDs in pre- and post-fatigue landings of a healthy athlete and an ACLR athlete.

2.6. Equivalent Damping Coefficient of Lower-Limb Muscles

To evaluate lower-limb muscles' equivalent damping and equivalent stiffness, a one degree-of-freedom (1-DOF) biomechanical model of the body is assumed [44,45]. In this model, which is shown in Figure (4), the body is modeled with an equivalent spring (K) and damper (C) for a muscle and an equivalent mass (M) for other parts. By assuming such a biomechanical model, the natural damping frequency of the muscle can be calculated as follows:

$$\omega_d = \omega_n \sqrt{1-\xi^2} \quad , \quad \left\{ \omega_n = \sqrt{\frac{K}{M}} \quad , \quad \xi = \frac{C}{2\sqrt{KM}} \right\} \tag{4}$$

Where $\omega_d$ indicates the natural damping frequency, $\omega_n$ is the undamped natural frequency and $\zeta$ is the damping ratio. $K$ indicates stiffness, $M$ is mass and $C$ is the damping coefficient. According to this equation, a decrease in the damping coefficient or an increase in the stiffness will consequently increase the natural damping frequency. Thus, the muscle's relative damping coefficient or relative stiffness can be inferred from the frequency of the sEMG signal. First, for each muscle, the overall mean of the MF in pre- and post-fatigue landings was calculated for each group (Table 1). Then the difference in the mean values between the two groups was calculated as a percentage (Mdiff%) of the healthy group's mean. These Mdiff% can be used in Eq.4 to determine the relative damping coefficient or relative stiffness.

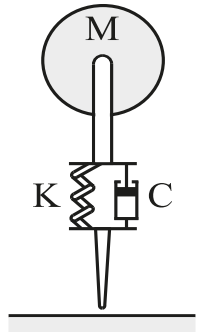

**Figure 4.** Assumed 1-DOF biomechanical model of the body for single-leg landing protocol.

2.7. Statistical Analysis

This cross-sectional study had 2 independent variables: group (dominant limb of the healthy group versus reconstructed limb of the ACLR group) and fatigue state (pre-fatigue versus post-fatigue). The dependent variables that were statistically analyzed in each muscle include the MF of the power spectrum density, the underneath volume of the PWVD in the pre-activity phase and in the reactive phase as a percentage of the total underneath volume. To detect the main effect of group and the main effect of fatigue state in each parameter, a 2×2 repeated measures analysis of variance (ANOVA) was used. Where necessary, paired samples t-tests were used to detect significant fatigue-induced variations within the groups, and one-way ANOVA was used to detect significant differences between the groups.

## 3. Results

The mean and the standard deviation (SD) of the calculated parameters, the results of one-way ANOVA for between-group analysis, and the results of paired samples t-tests for within-group analysis are reported in Table 1 and Table 2. For better interpretation, the results are illustrated as Mean±SD chart in Figure (5) as well.

The main effect of the group on the MF of the sEMG signal was significant in both hamstring muscles including LH (F=13.255, p=0.002) and MH (F=9.662, p=0.006). More precisely, the MF of the LH and MH muscles of the ACLR athletes was significantly higher than the healthy group in both pre- and post-fatigue landings. For all other muscles, the MF of the ACLR group was not statistically significantly different from that of the healthy group, either in pre-fatigue or post-fatigue landings. Also, as expected, the main effect of fatigue on the MF of the sEMG signal was significant in both hamstring muscles including LH (F=17.835, p<0.001) and MH (F=22.290, p<0.001), both gastrocnemius muscles including LG (F=62.743, p<0.001) and MG (F=42.302, p<.001), as well as VL (F=13.228, p=0.002) from the quadriceps muscles group. Specifically, fatigue reduced the MF of the above-mentioned muscles significantly in both the healthy and ACLR groups.

**Table 1.** Mean ± standard deviation (SD) of the median frequency (Hz) for each muscle's sEMG signal, calculated pre- and post-fatigue and overall mean (regardless of fatigue) for each group, within- and between-group significance levels as well as the difference in means between the two groups expressed as a relative percentage of the healthy group (Mdiff%).

| Muscle | Status/Significance | Healthy | ACLR | Between-Group P-Value | MDIFF% |
|---|---|---|---|---|---|
| VL | Pre-Fatigue | 56.4±13.2 | 59.6±13.4 | 0.584 | |
| | Post-Fatigue | 51.1±14.4 | 53.9±11.9 | 0.629 | |
| | Overall Mean | 53.8±13.7 | 56.7±12.7 | 0.463 | + 5.3 |
| | Within-Group P-Value | 0.027* | 0.028* | | |
| VM | Pre-Fatigue | 47.2±8.9 | 53.0±21.5 | 0.423 | |
| | Post-Fatigue | 45.2±8.0 | 51.0±13.7 | 0.241 | |
| | Overall Mean | 46.2±8.2 | 52±17.6 | 0.173 | + 11.7 |
| | Within-Group P-Value | 0.441 | 0.821 | | |
| LH | Pre-Fatigue | 51.8±9.4 | 63.8±11.5 | 0.015* | |
| | Post-Fatigue | 41.0±11.6 | 56.4±7.5 | 0.001* | |
| | Overall Mean | 46.4±11.6 | 60.1±10.2 | <0.001* | + 25.7 |
| | Within-Group P-Value | 0.012* | 0.014* | | |
| MH | Pre-Fatigue | 55.0±13.7 | 75.2±18.2 | 0.008* | |
| | Post-Fatigue | 44.9±9.2 | 56.3±12.8 | 0.028* | |
| | Overall Mean | 50±12.4 | 65.7±18.1 | 0.002* | + 27.2 |
| | Within-Group P-Value | 0.017* | 0.004* | | |
| LG | Pre-Fatigue | 74.4±11.7 | 83.1±9.4 | 0.072 | |
| | Post-Fatigue | 61.2±14.5 | 71.1±10.8 | 0.087 | |
| | Overall Mean | 67.8±14.4 | 77.1±11.6 | 0.055 | + 12.7 |
| | Within-Group P-Value | <0.001* | <0.001* | | |
| MG | Pre-Fatigue | 75.2±9.8 | 82.1±8.2 | 0.33 | |
| | Post-Fatigue | 64.1±12.8 | 73.0±8.0 | 0.066 | |
| | Overall Mean | 69.6±12.3 | 77.5±9.2 | 0.057 | + 10.7 |
| | Within-Group P-Value | <0.001* | 0.005* | | |

As mentioned in Table 1, based on the Mdiff%, the MF of the LH and the MH muscles in the ACLR group were 25.7% (p <0.001) and 27.2% (p = 0.002) higher than the control group respectively. This result indicates eighter higher muscular stiffness or lower damping coefficient in the LH and the MH muscles of the ACLR group compared to the healthy ones. Overly, the mean of all Mdiff% was 15.55%, which is an indicator of relative equivalent damping coefficient deficiency or excessive equivalent stiffness of all muscles of the ACLR group in comparison to the healthy group.

As mentioned in Table 2, in all muscles of both healthy and ACLR groups, the energy of the sEMG signals in the pre-fatigue landings is concentrated in the reactive phase. The effect of fatigue on the energy distributions of all muscles of the healthy group (except for MH) was to shift the PWVD energy to the pre-activity phase. Although this energy shift is also observed in the MH muscle of the healthy group, it was not statistically significant. However, muscular fatigue only caused a significant shift of the energy from the reactive phase to the pre-activity phase in the LH and LG muscles of the ACLR group. Also, in the lateral part of the hamstring muscle (LH) of the ACLR group, there is an inability to damp the vibrations. More specifically, in the LH muscle of the ACLR group, the volume under the PWVD in the reactive phase was significantly higher than for healthy athletes in pre-fatigue landings (p = 0.044), indicating an inability of this muscle to damp the energy produced in the muscle after ground contact in the ACLR group.

**Table 2.** Mean ± SD of the underneath volume of the PWVD in the pre-activity and the reactive phase for both pre- and post-fatigue landings of the healthy and ACLR athletes as a percentage of the total underneath volume of the PWVD and the significance levels for within- and between-group analysis.

| Muscle | Status/Significance | Pre-Activity Volume (% Total Volume) | | | Reactive Volume (% Total Volume) | | |
|---|---|---|---|---|---|---|---|
| | | Healthy | ACLR | Between-Group P-Value | Healthy | ACLR | Between-Group P-Value |
| VL | Pre-Fatigue | 15.0±10.1 | 22.3±17.8 | 0.252 | 84.9±10.1 | 77.7±17.8 | 0.252 |
| | Post-Fatigue | 27.9±15.3 | 26.0±14.6 | 0.762 | 72.0±15.3 | 73.9±14.6 | 0.762 |
| | Within-Group P-Value | <0.002* | 1.527 | | <0.002* | 1.527 | |
| VM | Pre-Fatigue | 14.9±10.0 | 25.0±25.1 | 0.229 | 85.0±10.0 | 74.9±25.1 | 0.229 |
| | Post-Fatigue | 25.4±18.9 | 29.2±17.3 | 0.621 | 74.6±18.9 | 70.7±17.3 | 0.621 |
| | Within-Group P-Value | 0.033* | 0.618 | | 0.033* | 0.618 | |
| LH | Pre-Fatigue | 38.9±9.6 | 30.4±8.9 | 0.044* | 61.0±9.6 | 69.5±8.9 | 0.044* |
| | Post-Fatigue | 48.5±11.1 | 44.2±9.7 | 0.354 | 51.4±11.1 | 55.7±9.7 | 0.354 |
| | Within-Group P-Value | 0.003* | 0.002* | | 0.003* | 0.002* | |
| MH | Pre-Fatigue | 30.6±18.3 | 35.9±19.5 | 0.519 | 69.3±18.3 | 64.0±19.6 | 0.519 |
| | Post-Fatigue | 33.6±21.0 | 35.2±22.3 | 0.867 | 66.3±21.0 | 64.7±22.3 | 0.867 |
| | Within-Group P-Value | 0.382 | 0.924 | | 0.382 | 0.924 | |
| LG | Pre-Fatigue | 39.6±22.1 | 36.2±10.7 | 0.654 | 60.3±22.1 | 63.7±10.7 | 0.654 |
| | Post-Fatigue | 59.1±21.2 | 48.2±12.4 | 0.157 | 40.8±21.2 | 51.7±12.4 | 0.157 |
| | Within-Group P-Value | 0.011* | 0.009* | | 0.011* | 0.009* | |
| MG | Pre-Fatigue | 38.0±28.7 | 55.1±21.1 | 0.128 | 61.9±28.7 | 44.8±21.1 | 0.128 |
| | Post-Fatigue | 66.4±28.8 | 53.5±28.5 | 0.305 | 33.6±28.8 | 46.5±28.5 | 0.305 |
| | Within-Group P-Value | 0.031* | 0.794 | | 0.031* | 0.794 | |

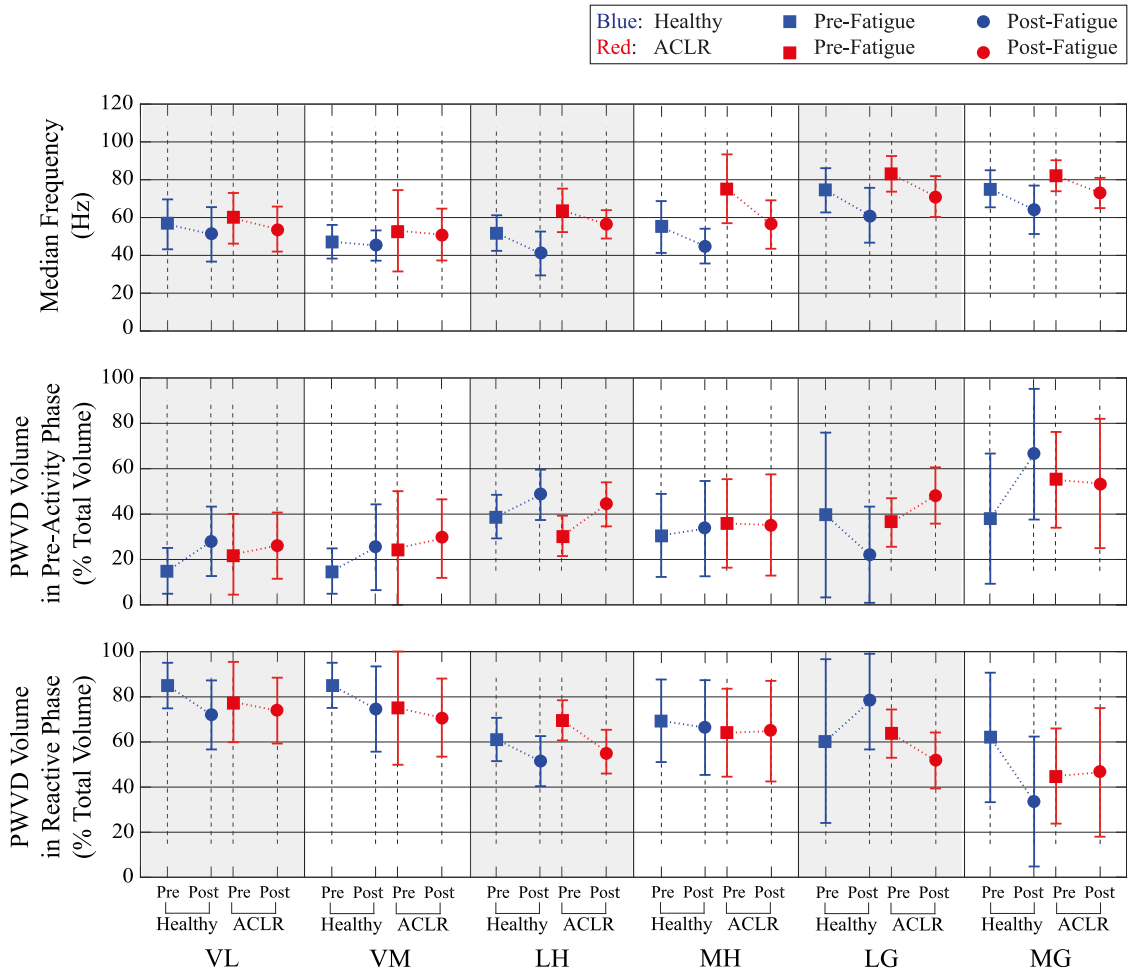

**Figure 5.** Mean±SD chart of the median frequency (Hz) of each muscle's sEMG signal, percentage of the underneath volume of the PWVD in pre-activity and reactive phase, in pre- and post-fatigue landings of the healthy and ACLR athletes.

## 4. Discussion

According to the literature, a decremental trend in MF was expected due to fatigue [37]. This is because, as a result of fatigue, the conduction velocity of the muscle fibers decreases, which subsequently leads to a decrease in the MF of the sEMG signal [34,36]. Consistent with this, fatigue significantly reduced the MF of the acquired sEMG signals in both healthy and ACLR groups in this study, as is illustrated in Figure (5) and noted in Table 1. The observation of the expected decrease in MF supports the validity of the implemented fatigue protocol. In addition, the MF of the sEMG signals of the ACLR athletes was significantly higher than the healthy athletes in both pre- and post-fatigue landings, which can be attributed to two different causes. A neuromuscular explanation for this phenomenon relates to alterations in the recruitment patterns of motor units within a muscle. A high recruitment and activation rate of fast motor units produces high-frequency components in the time-frequency analysis of electromyographic signals [46,47]. Fast intramuscular fibers are type II fibers that have a higher conduction velocity and greater force generation capacity than type I and, since the MF of the sEMG signal changes with variations in the conduction velocity of fibers, observation of a high MF could be the result of the overactivation of type II fibers [35-37,46,47]. It is notable that the more type II fibers a muscle has or activates, the more it will be prone to strain injuries [48]. Hence, a significantly higher MF of the sEMG signals from ACLR athletes' hamstring muscles compared with those of healthy athletes could reflect the predominant activation of type II fibers in their hamstrings. This predominant pattern, as a neuromuscular adaptation that develops 6 months postoperatively, could expose ACLR athletes to a greater risk of hamstring muscle strain injuries when returning to sport. Recently, it has been found that hamstring muscles of ACLR athletes generate greater forces than healthy subjects when single-leg landing [49], which is consistent with the neuromuscular explanation proposed to explain our results. As the result of the predominant activation of type II fibers in the hamstring muscles of ACLR athletes when single-leg landing, this muscle produces a higher MF in its sEMG signal and also greater muscular force. A biomechanical explanation of the higher MF of hamstring muscles' sEMG signals in the ACLR group, as mentioned in the results based on the Mdiff% values, reflects a significant damping deficiency or excessive stiffness in their hamstrings compared to the healthy athletes, at 6 months post-reconstruction. According to the literature, an increase in the stiffness of the knee joint or surrounding muscles after reconstruction surgery is a common complication that impairs knee function in ACLR patients [50,51]. This complication which is found using the proposed features in the hamstring muscles of the ACLR group of this study implies an inability to moderate forces, which could increase the probability of hamstring injuries caused by the large forces applied in high-impact sports.

The results of the PWVD analysis show that fatigue reduces the energy of the sEMG signal in the reactive phase and shifts the bulk of the energy to the pre-activity phase. Moreover, with the reduction of the MF due to fatigue, the energy distribution becomes concentrated mostly in lower frequencies. It has been shown that the spread of energy in the time-frequency plane of the PWVD indicates an inability of the muscle to damp the vibrations in the lower limb when making contact with the ground [29,30]. Conversely, the concentration of the energy in a small time-frequency interval indicates the muscle's ability to damp the vibrations [29,30]. This ability is quite evident in the pre-fatigue landings of healthy athletes, as illustrated in Figure (3). Also, as mentioned in the results section, the energy spread of the sEMG signal from the LH muscle in pre-fatigue landings of ACLR athletes was significantly higher than that of the healthy group. Thus, the PWVD time-frequency analysis in LH has demonstrated an inability of the hamstring muscles of ACLR athletes to damp vibrations. This result is in accordance with other studies reporting hamstrings strength deficit at 6 months post-reconstruction surgery of ACL [52]. Therefore, specialized pre-return-to-sport rehabilitation programs should be implemented to strengthen the hamstring muscles of ACLR athletes. By doing so, following passing the return-to-sport criteria, ACLR athletes could have a more successful return-to-sport with a low risk of ACL re-injury [53].

This study had some limitations. First, this study was conducted on male athletes and hence, the results may not be generalizable to ordinary people and females. Second, a single type of test (single-leg landing) was performed in this study. Hence, it is necessary to evaluate the proposed features when performing other plyometric exercises and dynamic tests in future studies. Third, this study was cross-sectional and hence, longitudinal cohort studies are necessary in future researches for long-term assessment of ACLR individuals using the proposed features. Lastly, this study was conducted in clinical environment. Nonetheless, the presented approach is not limited to clinical environment and the introduced sEMG-based features could also be incorporated in wearable tools instrumented with sEMG sensors for at-home assessments and remote monitoring that are rapidly being developed with advancement of technology [54].

## 5. Conclusion

In this research, three discriminative sEMG-based features are introduced to assess damping ability and interpret activation patterns in lower-limb muscles of ACLR athletes when performing a dynamic protocol. Assessing the proposed features revealed a significant damping deficiency in hamstring muscles of ACLR athletes at 6 months post-reconstruction surgery. It is clear that the quality of post-reconstruction and pre-return-to-sport assessments on ACLR subjects depends on the type and nature of the features extracted from the sEMG signals of the lower-limb muscles. Therefore, extraction of features with high differentiation ability and the ability to provide additional biomechanical information regarding the behavior of ACLR athletes' muscles should be in priority. This would facilitate the assessment process using non-invasive methods such as sEMG.

## Declaration of competing interest

The authors have no financial, professional, or personal interest of any nature of kind that could be construed as a potential conflict of interest or could be influencing the presented content.

## CRediT authorship contribution statement

**Mehran Hatamzadeh:** Conceptualization, Data curation, Formal analysis, Investigation, Methodology, Software, Writing - original draft, Writing - review & editing. **Ali Sharifnezhad:** Conceptualization, Methodology, Project administration, Supervision, Writing - review & editing. **Reza Hassannejad:** Conceptualization, Investigation, Methodology, Project administration, Supervision, Writing - review & editing. **Raphael Zory:** Supervision, Writing - review & editing.

## Acknowledgments

None.